\documentclass[12pt]{iopart}

\usepackage{iopams}
\usepackage[utf8]{inputenc}

\usepackage{mathtools}

\usepackage{float}
\usepackage{graphicx}
\usepackage{subcaption}
\usepackage{cite}
\usepackage{url}

\usepackage[numbers,sort&compress]{natbib}

\DeclarePairedDelimiter\ket{\lvert}{\rangle}
\DeclarePairedDelimiterX\braket[2]{\langle}{\rangle}{#1\,\delimsize\vert\,\mathopen{}#2}

\begin{document}

\title[Exploring Multiscale Quantum Media]{Exploring Multiscale Quantum Media: High-Precision Efficient Numerical Solution of the Fractional Schr\"odinger equation, Eigenfunctions with Physical Potentials, and Fractionally-Enhanced Quantum Tunneling}

\author{Joshua M. Lewis$^{1,2}$ and Lincoln D. Carr$^{1,2,3}$}

\address{$^{1}$Quantum Engineering Program, Colorado School of Mines, Golden, CO 80401, U.S.A.}
\address{$^{2}$Department of Physics, Colorado School of Mines, Golden, CO 80401, U.S.A.}
\address{$^{3}$Department of Applied Mathematics and Statistics, Colorado School of Mines, Golden, CO 80401, U.S.A.}
\ead{joshualewis@mines.edu}

\begin{abstract}
Fractional evolution equations lack generally accessible and well-converged codes excepting anomalous diffusion.  A particular equation of strong interest to the growing intersection of applied mathematics and quantum information science and technology is the fractional Schr\"odinger equation, which describes sub-and super-dispersive behavior of quantum wavefunctions induced by multiscale media.  We derive a computationally efficient sixth-order split-step numerical method to converge the eigenfunctions of the FSE to arbitrary numerical precision for arbitrary fractional order derivative.  We demonstrate applications of this code to machine precision for classic quantum problems such as the finite well and harmonic oscillator, which take surprising twists due to the non-local nature of the fractional derivative. For example, the evanescent wave tails in the finite well take a Mittag-Leffer-like form which decay much slower than the well-known exponential from integer-order derivative wave theories, enhancing penetration into the barrier and therefore quantum tunneling rates.  We call this effect \emph{fractionally enhanced quantum tunneling}.  This work includes an open source code for communities from quantum experimentalists to applied mathematicians to easily and efficiently explore the solutions of the fractional Schr\"odinger equation in a wide variety of practical potentials for potential realization in quantum tunneling enhancement and other quantum applications.\end{abstract}

%
%
%
%
%


\section{Fractional differential equations in context}

In recent decades, the field of fractional differential equations (FDEs) has  developed rapidly due to their wide applicability in physics, engineering, economics, and biology \cite{sun2018Applications, kulish2002application, wyss1986fractionalDiffusion, sokolov2002fractional, lai2016investigation,chen2022fractional, obembe2017fractional,zhang2021bio, ionescu2017role}.  The best known example of these is the anomalous diffusion equation, which has been experimentally verified to accurately describe diffusion on short time scales \cite{zhokh2017investigation, zhokh2017experimental}.  An outstanding question with FDEs is how to properly simulate them with well-converged and understood numerical methods so as to explore and predict such experimental scenarios.  For example, anomalous diffusion FDE simulators are used to predict diffusion of pollution in groundwater systems, where the fractional derivatives create a greatly increased probability of pollutants propagating to large distances \cite{chen2015transient}. Mathematically, this appears as a power-law probability distribution in place of the usual normal or Gaussian one, resulting in the distribution growing faster than the usual $t^{1/2}$ \cite{raghavan2011fractional}.  The anomalous diffusion FDE is a particular example of general reaction-diffusion FDEs which largely remain unsolved and unexplored.  In this Article, we focus on the fractional Schrödinger equation (FSE), connected to anomalous diffusion by a complex rotation of the time derivative from the real to the imaginary axis, just as for the integer-order case. 

Turning now to possible mathematical solution methods for FDEs, we first note that linear FDE methods such as the Laplace transform, Fourier transform, Mellin transform, Adomain decomposition, and separation of variables may be used effectively in only niche examples \cite{mainardi2007mellin, abdel2015natural}.  Unfortunately, in general the exact solutions to FDEs are intractable, due in part to the nonlocal nature of fractional derivatives. Such nonlocality can be observed in the many definitions of fractional derivatives. For example the Riesz fractional derivative is a particularly useful form for physics readers, since it generalizes from the Fourier transform $\partial/\partial x \to ik $ as follows: 
\begin{equation}
    (\partial^\alpha/\partial x^\alpha) f(x) \equiv \partial_x^\alpha f(x)\equiv \int dk \,e^{ikx} (-|k|^\alpha) \tilde{f}(k) \, .
\label{eqn:riesz}
\end{equation} In effect, this nonlocality creates an integrodifferential equation.  Thus, numerical methods are a vital component to understanding FDEs.  Direct implementation of fractional derivatives to discrete numerical schemes is subject to creating errors and inaccuracies just as with integer derivatives.  For example, famously the obvious direct discretization of the usual integer-order Schr\"odinger equation via finite-differencing does not conserve the normalization \cite{leforestier1991comparison}.  Many FDE methods developed so far, while useful for back of the envelope brainstorming, suffer from such issues.  Thus, in order to make accurate and reliable predictions for multiscale experiments, rigorous and careful development of such numerical methods is desirable.  The methods we develop in this Article will be generally useful toward this goal. As a clear demonstration and for context we will focus on solving the eigenfunction problem for the FSE in a variety of physically relevant potentials.  Eigenfunctions provide a complete basis for all solutions for linear FDEs just like integer order linear differential equations.

One tracing of the historical origin of the FSE can be pieced together as follows.  The original Schr\"odinger equation was reinterpreted as a path integral over Brownian motion by Feynman \cite{klauder2003feynman}.  In Brownian motion, the paths can be written as non-differentiable, self-similar curves with fractional dimensions that may differ from the Cartesian embedding in e.g. 3D. 
%
%
Laskin then constructed the FSE as a path integral over L\'evy flights \cite{laskin2000fractional}.  This construction is the natural generalization over paths with fractional dimension differing from Brownian motion. This change in the dispersive properties of the underlying equation creates derivatives of fractional order. The particular type of derivative that arises from this construction is the Riesz fractional derivative, Eq.~(\ref{eqn:riesz}). Although there are many useful definitions of fractional derivatives, such as the Caputo derivative used in some anomalous diffusion FDE simulation methods \cite{sun2010fractional}, we will focus on the Riesz derivative for the purposes of this Article.  

In the context of Laskin's many seminal contributions to the formulation and understanding of the FSE \cite{laskin2000fractional, laskin2000fractals, laskin2002fractional, laskin2010principles} we point out that he naturally used a common approach to solving the integer-order Schr\"odinger equation $\alpha=2$ for piece-wise potentials, namely matching the wavefunction and its first derivative at step-function boundaries.  This is a general method for linear ODEs commonly taught for a wide range of reaction-diffusion type equations in which the number of required matching conditions at each step-function is equal to one less than the order of the highest derivative.  Although not a universal method it applies in a wide range of cases.  The nonlocal nature of fractional derivatives is one definite exception to this approach, since regions bounding the step functions cannot be treated independently \cite{jeng2010nonlocality}.  This led to a misunderstanding of evanescent waves in Laskin's work \cite{laskin2000fractional} which we shall correct.  The essential elements of our numerical approach can be applied to a variety of reaction-diffusion and other ODE and even PDE contexts addressing decay of a solution in a barrier.  As particular applications to the FSE we will consider first the particle on a ring problem showing that the functional form is in this case independent of the order of the derivative, while the eigenvalue does depend on $\alpha$.   We then consider the quantum harmonic oscillator, where we show that decay into the barrier is greatly slowed for non-integer orders, enhancing the evanescent waves. Finally for the finite well and double well we show this enhanced  evanescent wave effect naturally leads to higher tunneling rates due to larger penetration into the barrier, which we call \emph{fractionally-enhanced quantum tunneling}, an effect one expects to observe in multiscale media in quantum systems.

\section{Overview of the Fractional Schr\"odinger Equation}

The one-dimensional FSE takes the following form: 
\begin{equation}
    i\hbar \partial_t \Psi = -C_{\alpha}\frac{\hbar^2}{2 m}\partial_x^{\alpha}\Psi + V(x)\Psi \,,
\end{equation}
where the Reisz fractional derivative of order $\alpha$ may be here defined more succinctly than Eq.~(\ref{eqn:riesz}) in terms of the Fourier transform $F$
\begin{equation}
    \partial_x^\alpha f(x) = \mathcal{F}^{-1}(-|k|^\alpha \mathcal{F}(f(x)))\, ,
\end{equation}
where $C_{\alpha}$ is a constant with physical units that correct those imposed by the fractional derivative. Clearly, the Reisz fractional derivative is a Fourier multiplier type operator and given a fractional order of $\alpha = 2$ we recover the ordinary second order derivative. 

To simplify our analysis we rescale our units such that the FSE simplifies to the following. 
\begin{equation}
    i\partial_t \Psi = -\frac{1}{2}\partial_x^{\alpha}\Psi + V(x)\Psi = \hat{H}\Psi\,.
\end{equation}
Given this construction we may define a time evolution operator by integrating out the first derivative with respect to time, 
\begin{equation}
    \Psi(x,\Delta t) = e^{-i\hat{H}\Delta t}\Psi(x,0)\,.
\end{equation}

The eigenfunction problem may then be defined as those functions for which application of our Hamiltonian applies only a scaling factor to our function:
\begin{equation}
    \hat{H}\phi_n(x) = E_n \phi_n(x) \,.
\end{equation}
The set of linearly independent eigenfunctions which fulfills this property defines a basis from which any solution may be decomposed \cite{akhiezer2013theory}. The goal of this paper is to construct an effective means by which we may numerically construct eigenfunctions within this set. The means by which we will construct these eigenfunctions is via imaginary time evolution through the Fourier split step method, as discussed in Sec.~\ref{sec:splitstep}. 


\section{The Mittag-Leffler Function}

Before discussing the Fourier split step method, a special function with many connections to fractional derivatives must be briefly discussed. The Mittag-Leffler function arises naturally in the solution of many fractional order integral equations or fractional order differential equations and has diverse application in the study of fractional relaxation and fractional diffusion \cite{kilbas2004generalized, rahman2017extended}. Although the eigenfunctions generated to solve the fractional Schrodinger equation are not quantitatively described by the Mittag-Leffler function, certain behavior attributed to the Mittag-Leffler function and its family of functions appears when examining the eigenstates, especially the decay patterns in the classically forbidden regions of the eigenfunctions. 

\begin{figure}[H]
\centering
\begin{subfigure}{.5\textwidth}
  \centering
  \includegraphics[width=.99\linewidth]{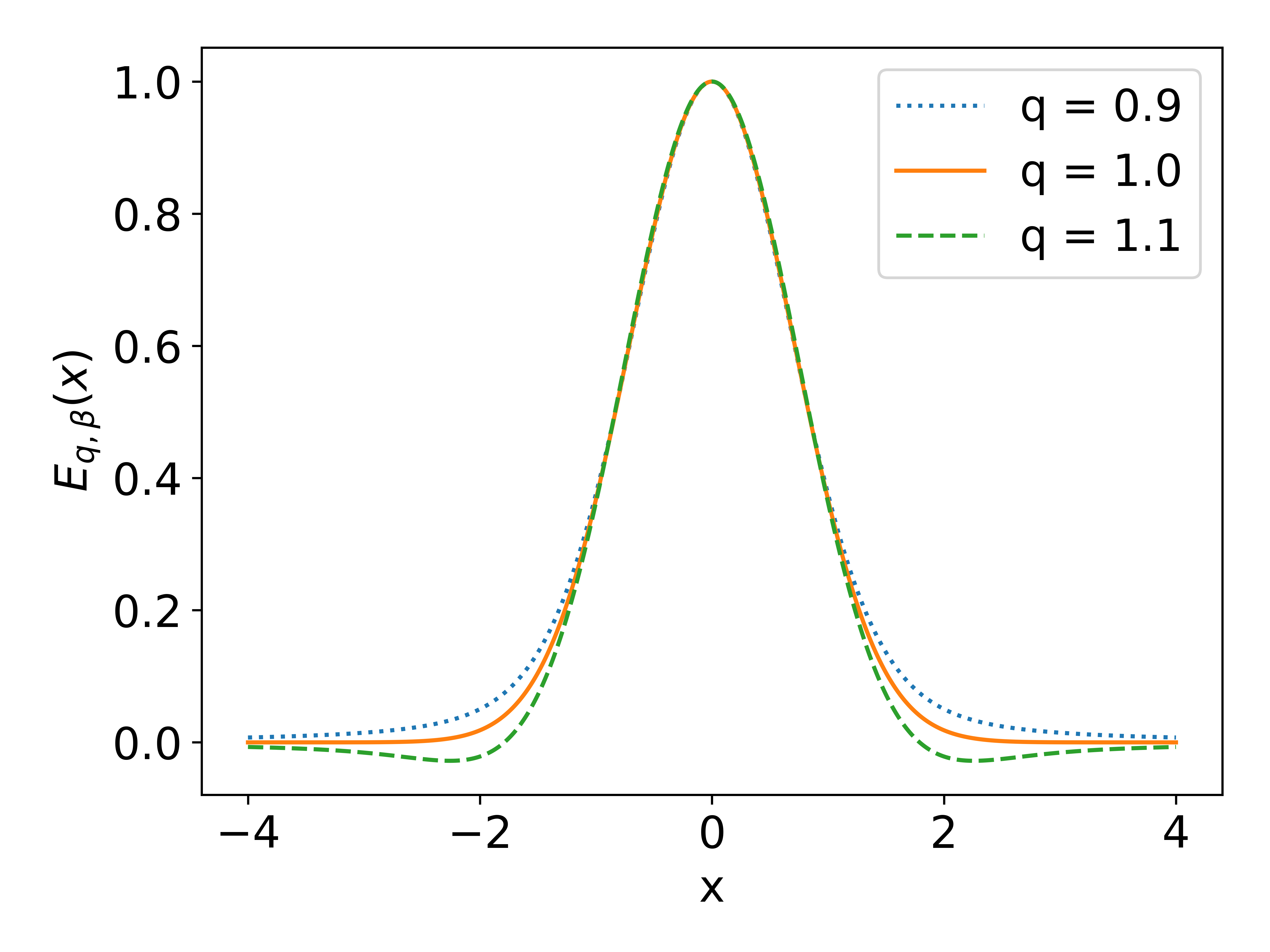}
  \caption{}
  \label{fig:sub1}
\end{subfigure}%
\begin{subfigure}{.5\textwidth}
  \centering
  \includegraphics[width=.99\linewidth]{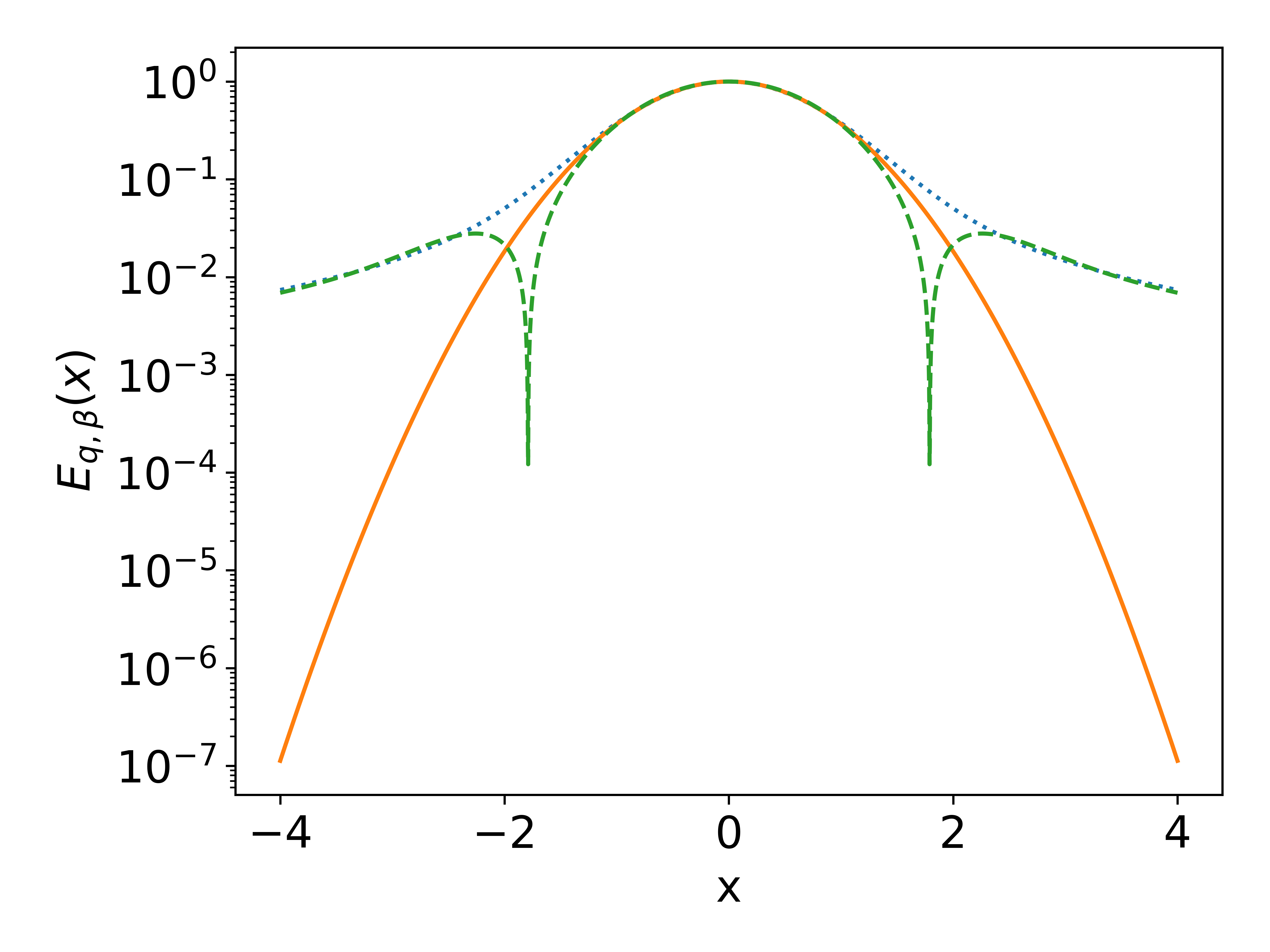}
  \caption{}
  \label{fig:sub2}
\end{subfigure}
\caption{\emph{Comparison of the Mittag-Leffler function to the Gaussian.} (a): Core behavior of the Mittag-Leffler function. (b): Logarithmic scale emphasizing the tails.}
\label{fig:MLBehavior}
\end{figure}

The Mittag-Leffler function can be understood as a type of generalization to the exponential function and is defined with one or two parameters via the following sums:
\begin{equation}
    E_{q} (x) = \sum_{k = 0}^{\infty}\frac{x^k}{\Gamma(q k + 1)}\,,
\end{equation}
\begin{equation}
    E_{q,\beta} (x) = \sum_{k = 0}^{\infty}\frac{x^k}{\Gamma(q k + \beta)}\,.
\end{equation}
Where the exponential function is the eigenfunction of the derivative operator, the Mittag-Leffler function is the eigenfunction of the Caputo fractional derivative, a fractional derivative connected to the fractional Rietz operator we study in this article \cite{peng2010note}. As a practical example of the type of behavior that the Mittag-Leffler function portrays, Fig.~\ref{fig:MLBehavior} compares a Gaussian and its Mittag-Leffler analog.

Clearly the Gaussian function is simply the Mittag-Leffler function given that $q = 1$ with the parameter $-x^2$. Additionally, close to the origin the functions are similar with slight divergences of slower or faster decay. However, the important piece is in the tails of Fig.~\ref{fig:MLBehavior}(b). Away from the central curve, the tails present a stark difference for non-integer $q$. Examining the case of $q = 0.9$, they decay significantly slower than the Gaussian case and in fact the decay appears to slow down as the tail extends further. Furthermore, for $q = 1.1$ we observe the main difference is the appearance of a node before it again slowly decays towards zero. In both cases we find the tails decay significantly slower for non-integer $\alpha$ as compared to the integer case.
%

To summarize, where typically we would observe exponentially decaying tails in integer order eigenfunctions, what we find in our generated eigenfunctions with the FSE is Mittag-Leffler-like decaying tails. Although not exactly the same, certain stark features very reminiscent of Mittag-Leffler functions and related functions may be observed.

\section{Fourier Split Step: The Fractional Adaptation}
\label{sec:splitstep}

%
%
%

Recall that imaginary time evolution takes the form
\begin{equation}
    f_0(x,-it) = e^{-\hat{H}t}f_0(x) = \sum_n a_n e^{-E_n t} \psi_n(x)
\end{equation}
with $\psi_n(x)$ the eigenfunctions of a Hamiltonian and $a_n$ the initial weights.  All higher energy components decay exponentially more rapidly in imaginary time, resulting in the ground state.  To obtain higher eigenfunctions, one simply subtracts lower eigenfunctions at every renormalization step.
%
%
%
A modification to this method must be used to treat nearly degenerate or degenerate eigenstates, as one encounters e.g. in the double well potential. 
For the double well it is sufficient to constrain the parity to be even in order to obtain the ground state, and use alternating parity constraints for excited states. 


The difficulty then is not how to generate the eigenfunctions, but how to evolve in time given a fractional derivative in the Hamiltonian. Although the Reisz fractional derivative is quite complicated and difficult to apply in position space, its application in momentum space is straightforward as per Eq.~(\ref{eqn:riesz}).  Thus we have a portion of our Hamiltonian which is easy to apply in position space and difficult to apply in momentum space, the potential; and a portion which is easy to apply in momentum space and difficult to apply in position space, the fractional derivative. The Fourier split step method is a well-known solution to this circumstance, which we here write out plainly for the fractional case.    

\vspace{10pt}

Given that we may split our Hamiltonian into these components where $\hat{M}$ is the operator that is easy to apply in the momentum domain, and $\hat{P}$ is the operator that is easy to apply in the position domain, the problem is that we may not immediately separate these operators within the exponential function as these operators in general do not commute. However, by ignoring this and splitting them regardless we may construct an approximation that is 1st order in terms of the time step, $\Delta t$,
\begin{equation}
    \hat{H} = \hat{M} + \hat{P}\,,
\end{equation}
\begin{equation}
    e^{-\hat{H}\Delta t} = e^{-\hat{M}\Delta t}e^{-\hat{P}\Delta t} + \mathcal{O}(\Delta t)\,.
\end{equation}
As per the usual split step method, we Fourier transform the state before applying $\hat{M}$, then inverse Fourier transform, apply $\hat{P}$, and obtain the first order approximation. Given that our operators have the following form the complete procedure is as follows:
\begin{align}
    \hat{P} = V(x) && \hat{M} = \frac{1}{2}|k|^\alpha\,,
\end{align}
\begin{equation}
    \psi(x,-i\Delta t) =  \mathcal{F}^{-1}\left(e^{-\frac{1}{2}|k|^\alpha\Delta t}\mathcal{F}\left( e^{-V(x)\Delta t}\psi(x,0)\right)\right) + \mathcal{O}(\Delta t)\,.
\end{equation}

We may then increase the accuracy of this scheme by adding more steps and changing the constants on the operators such that the exponential is applied more accurately. A general construction of a splitting scheme is
\begin{equation}
    e^{\hat{A} + \hat{B}} \approx \prod_{k=1}^n e^{a_k\hat{A}}e^{b_k \hat{B}}\,.
    \label{eqn:splitting}
\end{equation}
An example of a different scheme that may be recovered via this procedure is the second order Suzuki-Trotter approximation, which is also a second order approximation in time:
\begin{equation}
    e^{(\hat{A} + \hat{B})\Delta t} = e^{\frac{1}{2}\hat{A}\Delta t}e^{\hat{B}\Delta t}e^{\frac{1}{2}\hat{A}\Delta t} + \mathcal{O}(\Delta t^2)\,.
\end{equation}

However, there is a significant subtlety here -- the condition on which constants $a_n$ and $b_n$ are allowed in order to keep the scheme unconditionally stable for the case of imaginary time evolution.  All constants $a_n$ and $b_n$ must be positive for all $n$. Unlike the real time evolution case where the applied complex exponential functions are rotations in complex space, imaginary time evolution has either a decaying or growing exponential. If $a_n$ or $b_n$ is negative then higher energy eigenfunctions grow significantly faster during specific steps than the lower energy eigenfunctions that we wish to isolate. Furthermore, machine error will produce contributions of very high order eigenfunctions that will grow significantly faster and lead to numerical instability. This problem is the same as that of evolving the diffusion equation backwards in time.  The problem that then arises is that splitting methods with approximation orders greater than 2nd order must necessarily have negative splitting coefficients \cite{blanes2005necessity}.
We can fix this is by allowing our splitting coefficients to not only be real valued but also complex valued. Given this adjustment it is easy to show that higher-order splitting schemes become available at the cost of the additional overhead of using complex numbers. Given this additional freedom, we choose to use a sixth order scheme that takes 8 steps. The coefficients from Eq.~(\ref{eqn:splitting}) are found in Table~\ref{table:splitting}.

\begin{table}[h]
\centering
\begin{tabular}{|l|l|l|}
\hline
Step & $a_n$                                                                                    & $b_n$                                                                                    \\ \hline
1    & \begin{tabular}[c]{@{}l@{}}0.0584500187773306\\ +0.0217141273080301i\end{tabular} & \begin{tabular}[c]{@{}l@{}}0.116900037554661\\ +0.0434282546160603i\end{tabular}  \\ \hline
2    & \begin{tabular}[c]{@{}l@{}}0.123229569418374\\ -0.0402806787860161i\end{tabular}  & \begin{tabular}[c]{@{}l@{}}0.129559101282088\\ -0.123989612188092i\end{tabular}   \\ \hline
3    & \begin{tabular}[c]{@{}l@{}}0.158045797047111\\ -0.0604410907390099i\end{tabular}  & \begin{tabular}[c]{@{}l@{}}0.186532492812133\\ +0.00310743071007267i\end{tabular} \\ \hline
4    & \begin{tabular}[c]{@{}l@{}}0.160274614757183\\ +0.0790076422169959i\end{tabular}  & \begin{tabular}[c]{@{}l@{}}0.134016736702233\\ +0.154907853723919i\end{tabular}   \\ \hline
5    & $a_4$                                                                                   & $b_3$                                                                                    \\ \hline
6    & $a_3$                                                                                   & $b_2$                                                                                    \\ \hline
7    & $a_2$                                                                                    & $b_1$                                                                                    \\ \hline
8    & $a_1$                                                                                    & 0                                                                                       \\ \hline
\end{tabular}
\caption{The splitting coefficients used in the FSE algorithm to obtain a sixth order approximation.   \label{table:splitting}}
\end{table}

To summarize, our FSE code is based on a sixth-order split-step method within the framework of the Suzuki-Trotter approximation. This approach facilitates the efficient application of the Riesz fractional derivative by leveraging the Fast Fourier Transform (FFT). A notable aspect of using FFT in our simulations is the implicit assumption of periodic boundary conditions. However, for scenarios involving bound states, open boundary conditions can be effectively emulated by simply expanding the system size. This expansion ensures that the boundary effects do not significantly influence the dynamics within the region of interest.

In this context, when evolving the Schrödinger equation in imaginary time, which corresponds to the real-time evolution of a diffusion equation with an added potential term, our approach employs the Fourier split-step method. This method, by its very nature, is unconditionally stable when applied to linear, time-independent problems. Its stability stems from the precise handling of the linear part of the equation in Fourier space, complemented by the treatment of the nonlinear or potential-dependent part in real space within each discrete time step. This characteristic of unconditional stability is particularly advantageous for simulations aiming to reach the ground state, as does not induce a restriction of the values of $\Delta t$ and $\Delta x$, this then permits the use of larger time steps without risking numerical instability, thereby facilitating efficient convergence over extended imaginary time scales.

However, the integration of certain types of potentials into the simulation imposes additional considerations related to spatial and frequency resolutions. Specifically, potentials characterized by large magnitudes, abrupt spatial variations, or significant nonlinearities necessitate a meticulous choice of spatial grid size ($\Delta x$) and time step ($\Delta t$). The presence of steep potential gradients calls for a denser spatial grid to accurately capture these variations, while high-frequency components induced by the potential require careful management in the frequency domain to prevent aliasing effects. Therefore, while the primary stability concern of the Fourier split-step method in the temporal domain is addressed, the introduction of complicated potentials demands a nuanced approach in spatial discretization and frequency domain analysis. This is essential to ensure the accuracy and reliability of the simulation results, particularly in quantum systems where the potential landscape critically influences the system dynamics.

In this study, an extensive convergence analysis was conducted to identify the optimal grid sizes that effectively minimize finite grid errors below the threshold relevant to our function analyses. For the entirety of the results presented in this paper, we standardized the grid size and time step at $\Delta x = 10^{-2}$ and $\Delta t = 10^{-2}$. An exception is made in the case of the finite well problem, where an additional step is incorporated to accurately address the sharp potential variations. Details of such adjustments are elaborated in the corresponding sections of the paper.

In line with our commitment to open science and reproducibility, the complete source code for our Fourier Split-Step simulations, as described in this paper, is made publicly available. Interested researchers can access, review, and utilize the code through our GitHub repository, titled "FractionalEigenstate." This repository includes detailed documentation and example scripts, providing a comprehensive guide for users to adapt and apply the code to their specific research needs. For access to the repository, please refer to our citation below \cite{FractionalEigenstate2024}.

\section{Demonstration of FSE Code Precision: Particle on a Ring}


To examine the effectiveness of this algorithm on fractional order quantum systems we first investigate one of the very few systems in which the analytical energies and eigenstates are well known, the particle on a ring system. This will enable a careful evaluation of our code's precision.

We define the particle on a ring to be a constant potential, here chosen to be zero, under periodic boundary conditions, with domain $x \in [-1, 1]$. The Hamiltonian is thus
\begin{equation}
    \hat{H} = -\frac{1}{2}\partial_x^{\alpha}.
\end{equation}

\begin{figure}[H]
    \centering
    \includegraphics[width=100mm,scale=0.5]{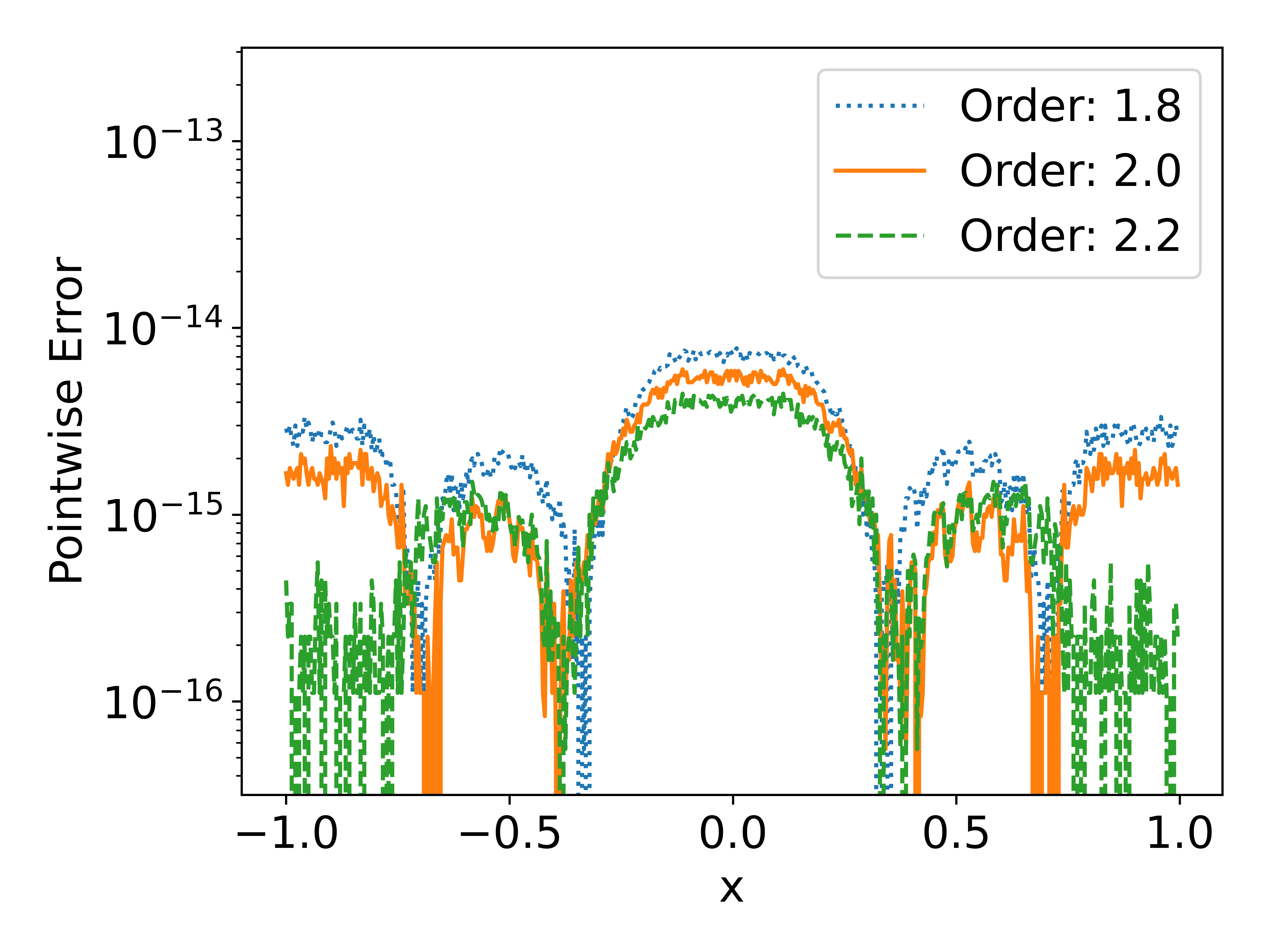}
    \caption{\emph{Point-wise error associated with the 1st excited state of the particle on a ring system across fractional orders.}}
    \label{fig:ringPointwiseError}
\end{figure}

Given that the Reisz fractional derivative is a Fourier multiplier similar to integer derivatives, the eigenfunctions of such an operator by itself are plane waves. As we cast our eigenfunctions to the real plane to ensure uniqueness these eigenfunctions are then just sines and cosines similar to the particle on a ring problem with the second derivative. Because of this, the eigenfunctions of this system are invariant of fractional order. The only aspect of the eigen-problem that does change is the energy, which we can obtain by imaginary time evolution, as per Sec.~\ref{sec:splitstep}, as well as analytically for this special case:
\begin{equation}
    \partial_x^{\alpha}\sin(kx) = -|k|^{\alpha}\sin(kx)\,,
\end{equation}
\begin{equation}
    E_n =\frac{1}{2}\left(\pi \lceil n/2 \rceil \right)^{\alpha}\,,
\end{equation}
where $\lceil * \rceil$ denotes the ceiling function. This is due to the degeneracy of the energies within the system. The reader will note that the usual integer order expression for the energy is recovered when $\alpha = 2$.
The eigenstates of the system are given by
\begin{equation}
\psi_n(x) =
    \begin{cases}
        \cos{\pi nx}       & \text{n even} \\
        \sin{\pi nx}       & \text{n odd} \,,
    \end{cases}
\end{equation}
independent of fractional order.
To ascertain the effectiveness of this algorithm, we first compare the generated first excited state for the particle on a ring system to its analytical eigenstate across a few fractional orders. For this examination, we define our grid in the domain [-1,1] with 480 evenly spaced points, a choice strategically made to enable the representation of the 39th excited state within the constraints of the Nyquist sampling theorem. The results of this analysis are presented in Figure \ref{fig:ringPointwiseError}. Interestingly higher order systems tend to converge faster and yield more accurate results than their lower order counterparts. Although not universally applicable, in many systems, an increase in the fractional order leads to a greater energy spacing between eigenstates. This increase in energy spacing inherently accelerates the convergence process via imaginary time evolution.

We now focus on evaluating the maximum error in the pointwise approximation of each eigenstate, with a particular interest in how this error varies in relation to the ordinal number of the eigenstate. To conduct this analysis, we opt for a discretization of the domain into 480 uniformly distributed points. The rationale behind choosing this specific number of points is grounded in the principles of the Nyquist sampling theorem. According to this theorem, to accurately represent a signal with a frequency of $f = \pi n$, a minimum of $4\pi n + 1$ points is required. Hence, for effectively capturing the nuances of the 39th excited state, which is represented by the function $\cos{39 \pi x}$, a minimum of 479 points is necessary. Therefore, our selection of 480 points ensures that we adequately sample the highest frequency state in our analysis, minimizing the potential for aliasing errors.

\begin{figure}[H]
    \centering
    \includegraphics[width=100mm,scale=0.5]{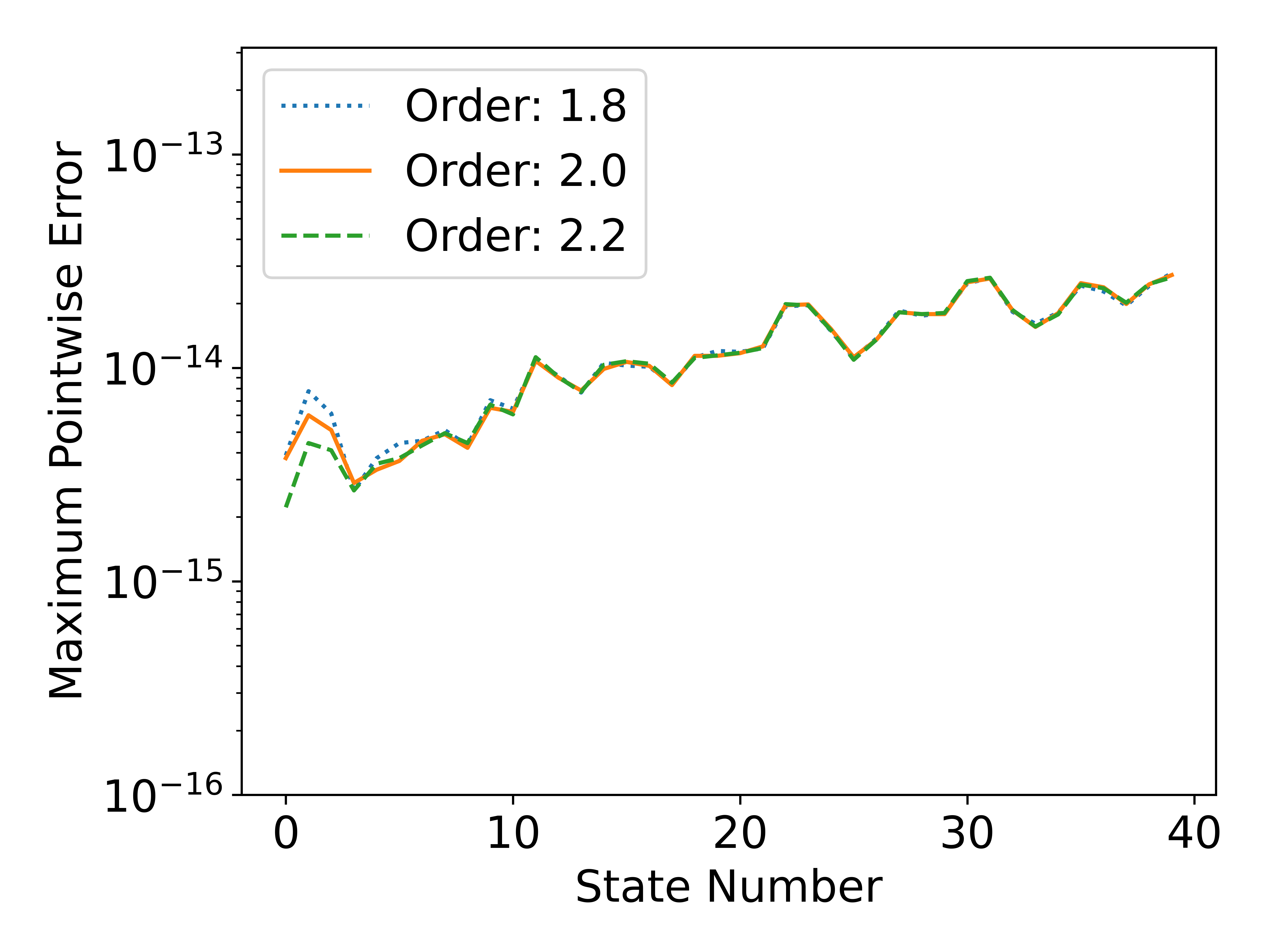}
    \caption{\emph{Maximum point-wise error across many eigenstates of the particle on a ring system for a few fractional orders.}}
    \label{fig:ringMaxPointwiseError}
\end{figure}

Finally we may compare the calculated energies from our generated eigenstates to the analytical energies for this system.

\begin{figure}[H]
    \centering
    \includegraphics[width=100mm,scale=1]{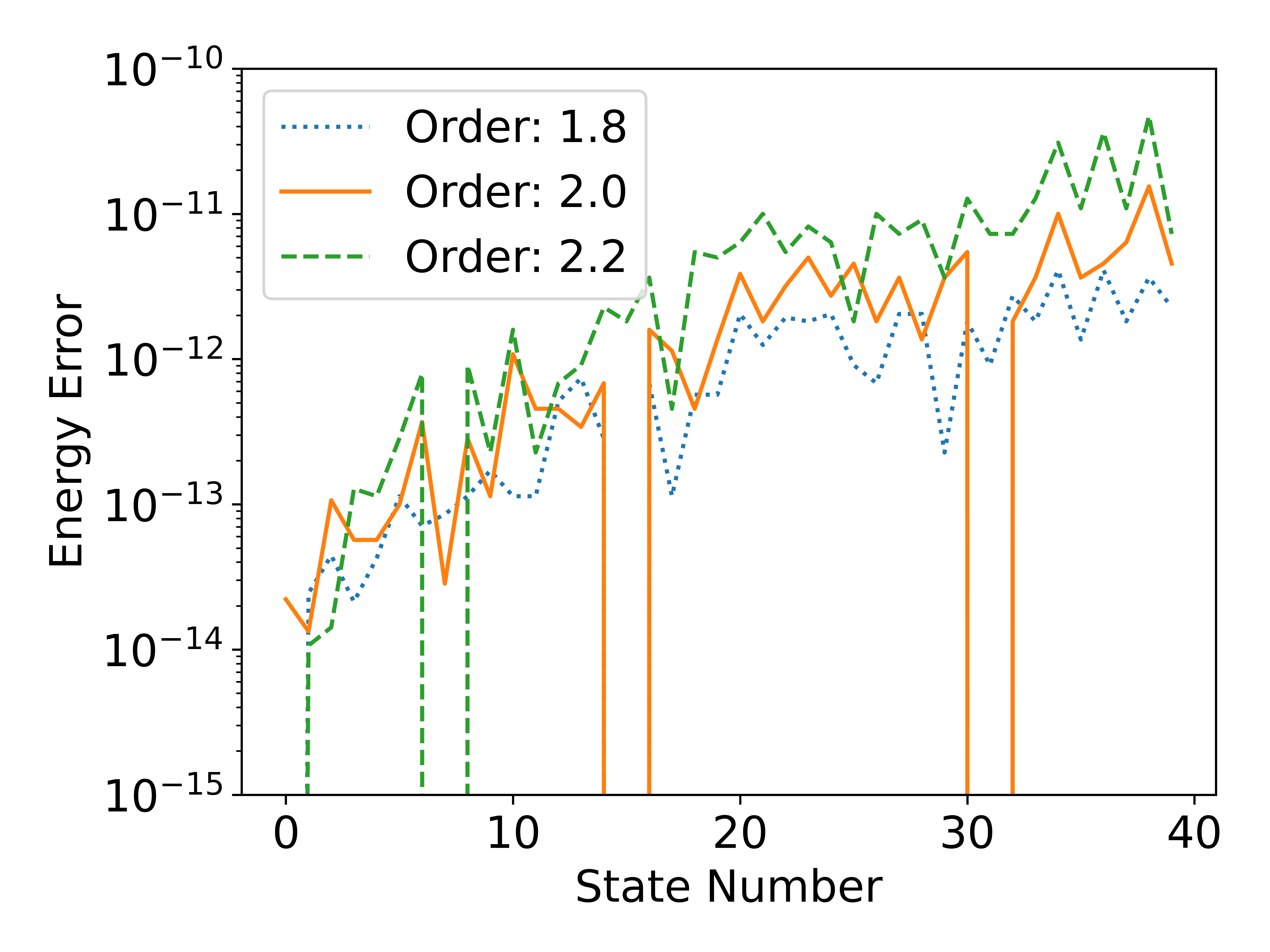}
    \caption{\emph{Error in generated energies when compared to analytical energy.}}
    \label{fig:my_label}
\end{figure}

\section{Fractional Quantum Mechanics Eigensystems}

Having established the efficacy of our code for the eigenvalue problem, we proceed to treat three classic quantum problems of physical interest for fractional systems: the quantum harmonic oscillator, the finite well, and the double well. We show how the new parameter $\alpha$ in the FSE creates new features. The double well in particular contains the essential elements of quantum tunneling between discrete energy levels, and has some quite surprising new aspects, which we call fractionally-enhanced quantum tunneling.


\subsection{Quantum Harmonic Oscillator}

The quantum harmonic oscillator problem is a somewhat famous problem within the study of quantum mechanics as the potential in the system well approximates many physically important systems and eigenstates produced lead to the construction of the coherent state and describe many important physical phenomena. The system be defined with the following Hamiltonian.

\begin{equation}
    \hat{H} = -\frac{1}{2}\partial_x^{\alpha} + \frac{1}{2}x^2
\end{equation}

\vspace{15pt}

\begin{figure}[H]
    \begin{subfigure}{.5\textwidth}
        \centering
        \includegraphics[width=1\linewidth]{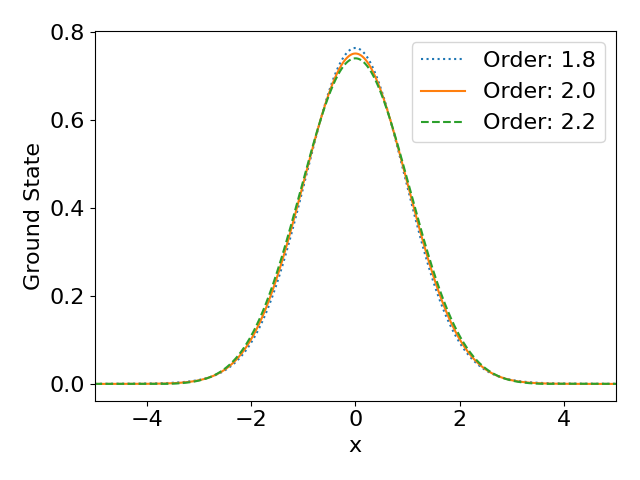}
        \caption{\emph{}}
        \label{fig:QHOPlot}
    \end{subfigure}%
    \begin{subfigure}{.5\textwidth}
        \centering
        \includegraphics[width=1\linewidth]{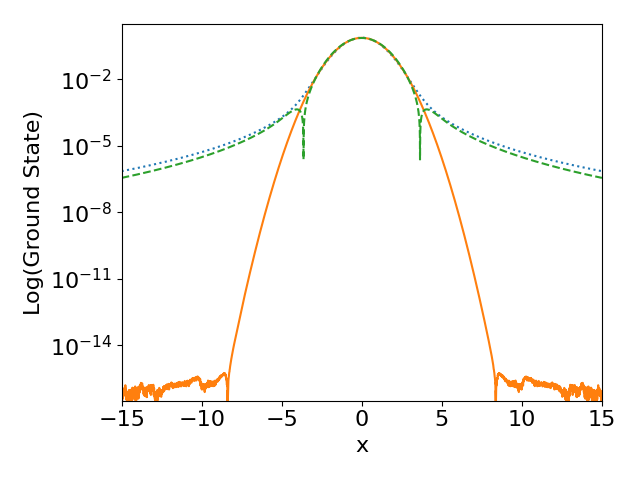}
        \caption{\emph{}}
        \label{fig:QHOLOGPlot}
    \end{subfigure}
    \begin{subfigure}{\textwidth}
        \centering
        \includegraphics[width=0.5\linewidth]{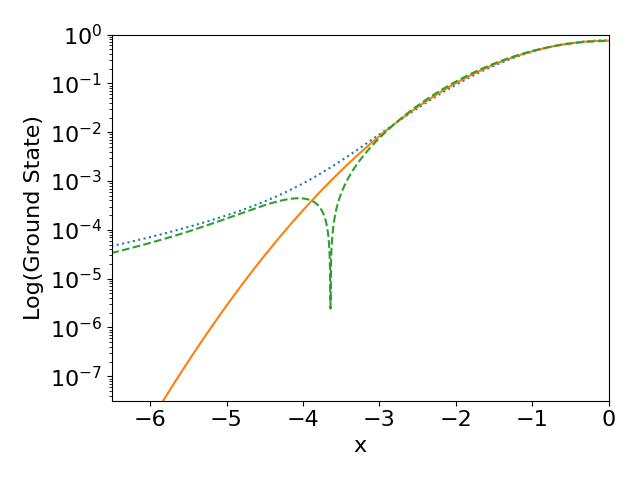}
        \caption{\emph{}}
        \label{fig:QHOLOGPlotZoomed}
    \end{subfigure}
    \caption{\emph{Eigenfunctions of the quantum harmonic oscillator potential across fractional orders: (a) Ground states depicted across fractional orders less than, equal to, and greater than 2. (b) Logarithmic view emphasizing tail visibility. (c) Close-up at the transition into the classically forbidden region, highlighting the distinct Mittag-Leffler-like tails and node in the higher order eigenstate.}}
    \label{fig:QHOPlots}
\end{figure}

Contrary to the prior system the eigenfunctions are no longer invariant of fractional order. We compare the ground state in Fig. \ref{fig:QHOPlot} across different orders to see this. While these different ground states appear to be of the same structure with slight variations, the greatest difference is hidden from sight in this plot. To uncover the main difference between these ground states we will analyze the log plot in Fig. \ref{fig:QHOLOGPlot}. 

The tails of the ground state are greatly altered by changing the fractional order. We see that the exponential decay from the integer order case has been replaced by Mittag-Leffler like decay in the tails. The node in the order 2.2 case is a clear hint of the Mittag-Leffler like nature of this eigenstate. One may refer back to Fig. \ref{fig:MLBehavior} to see the very similar behavior of the tails in the classically forbidden region.

We may then recover the energies associated with each eigenfunction by again evolving the state for a short time in imaginary time and extracting the energy from the decay rate.

\begin{table}[H]
\centering
    \begin{tabular}{|l|l|l|l|}
    \hline
        Eigenfunction & $\alpha = 1.8$ & $\alpha = 2.0$ & $\alpha = 2.2$ \\ \hline
        0   & 0.4994984133  & 0.4999999999  & 0.5012687387                \\ \hline
        1   & 1.4452024402  & 1.5000000003  & 1.5508889832                \\ \hline
        2   & 2.3415247398  & 2.4999999995  & 2.6507789545                \\ \hline
        3   & 3.2229118004  & 3.4999999991  & 3.7689276297                \\ \hline
        4   & 4.0884863294  & 4.4999999983  & 4.9044885301                \\ \hline
    \end{tabular}
\caption{\emph{Energies for the first five eigenfunctions associated with the quantum harmonic oscillator of orders 1.8, 2.0, and 2.2}}
\end{table}

One may notice that the linear eigenspectrum afforded to the system with integer order is entirely lost when the fractional order is changed. Thus, changing the fractional order of the quantum harmonic oscillator results in an eigenspectrum that does not have a revival time nor does it have a similar coherent state. We may then plot the spectrum of eigenvalues across many orders to see a clearer trend of how the eigenspectrum changes upon a changing fractional order. 

\begin{figure}[H]
    \centering
    \includegraphics[width=110mm,scale=1.0]{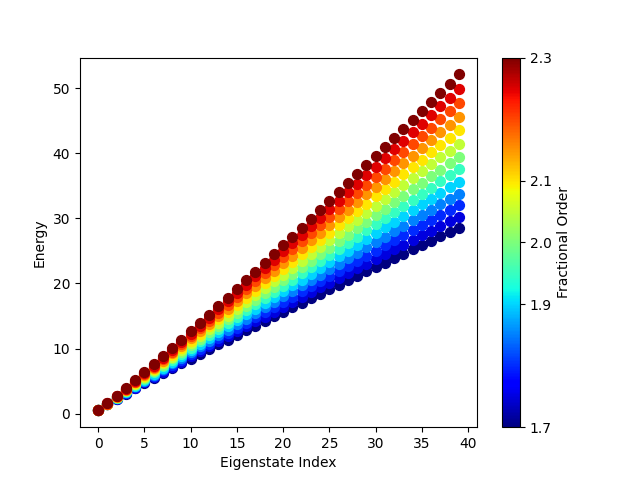}
    \caption{\emph{Spectrum of energies associated with eigenfunctions of the fractional quantum harmonic oscillator across many orders}}
    \label{fig:my_label}
\end{figure}

\subsection{Finite Well}

To examine the tails in the classically forbidden region and the compaction or expansion associated with the spectrum of eigenenergies upon a changing fractional order, we now investigate the finite well problem which may be defined with the following Hamiltonian.

\begin{equation}
    \hat{H} = -\frac{1}{2}\partial_x^{\alpha} + V(x)
\end{equation}

\begin{equation}
V(x)=
    \begin{cases}
        V_0 & \text{otherwise} \\
        0 & \text{if } |x| < 1 
    \end{cases}
\end{equation}

We will choose to set $V_0$ to 100 in this analysis such that a sufficient number of bound eigenstates may be generated within the well. We observe the ground state across three different fractional orders such that adequate comparisons may be made.

\begin{figure}[H]
    \begin{subfigure}{.5\textwidth}
        \centering
        \includegraphics[width=1\linewidth]{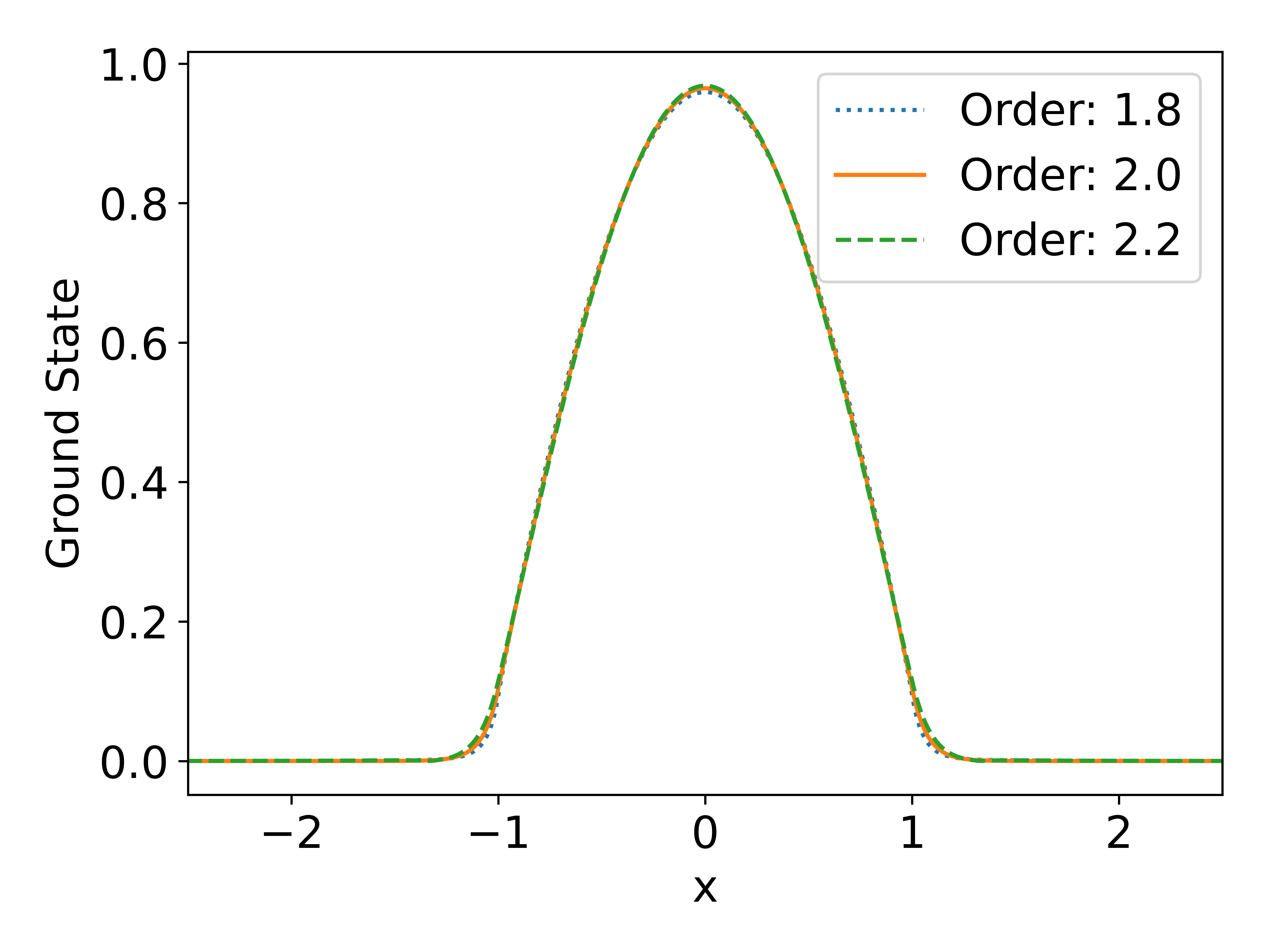}
        \caption{\emph{}}
        \label{fig:finiteWell}
    \end{subfigure}%
    \begin{subfigure}{.5\textwidth}
        \centering
        \includegraphics[width=1\linewidth]{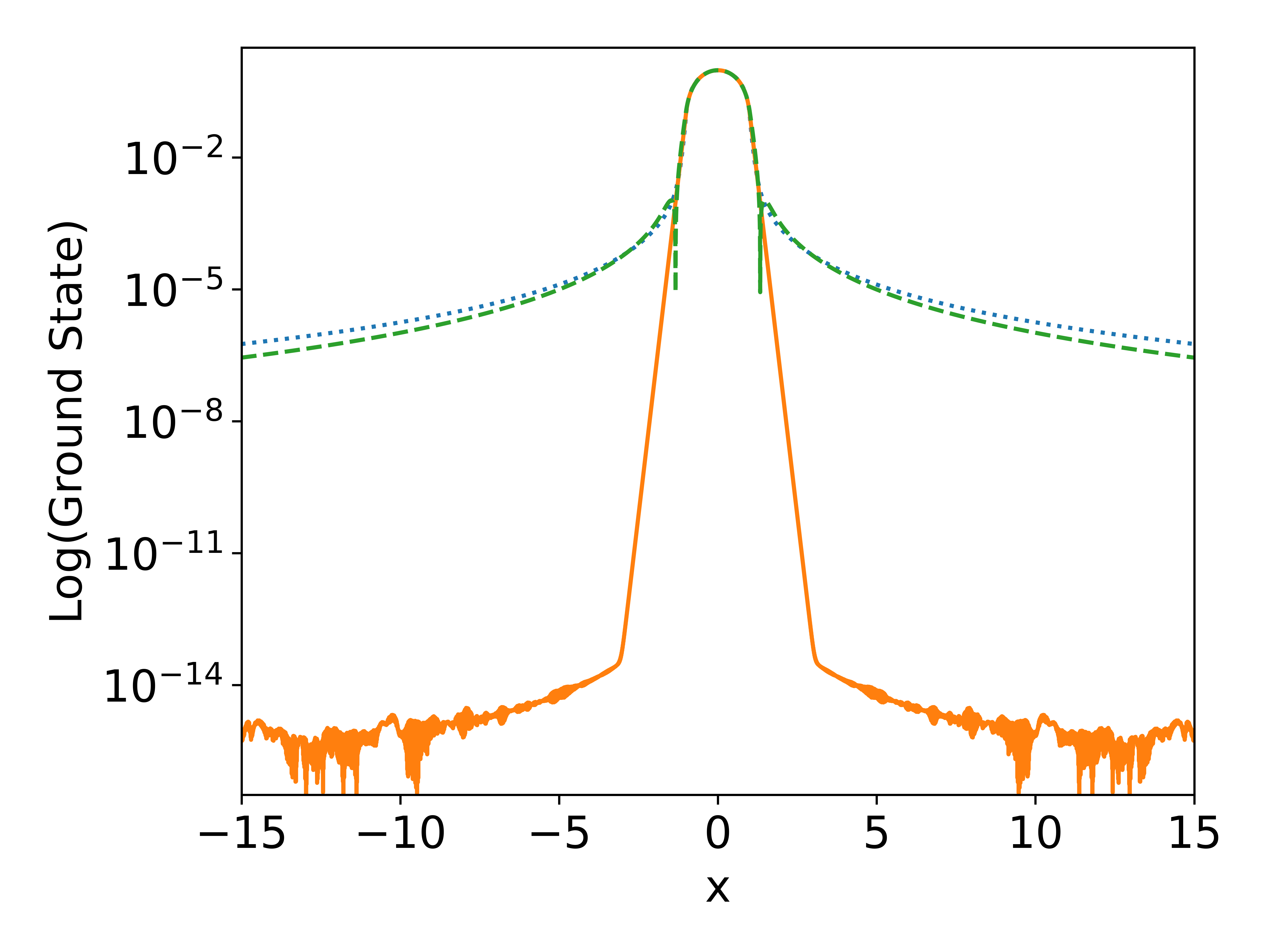}
        \caption{\emph{}}
        \label{fig:finiteWellLog}
    \end{subfigure}
    \begin{subfigure}{\textwidth}
        \centering
        \includegraphics[width=0.5\linewidth]{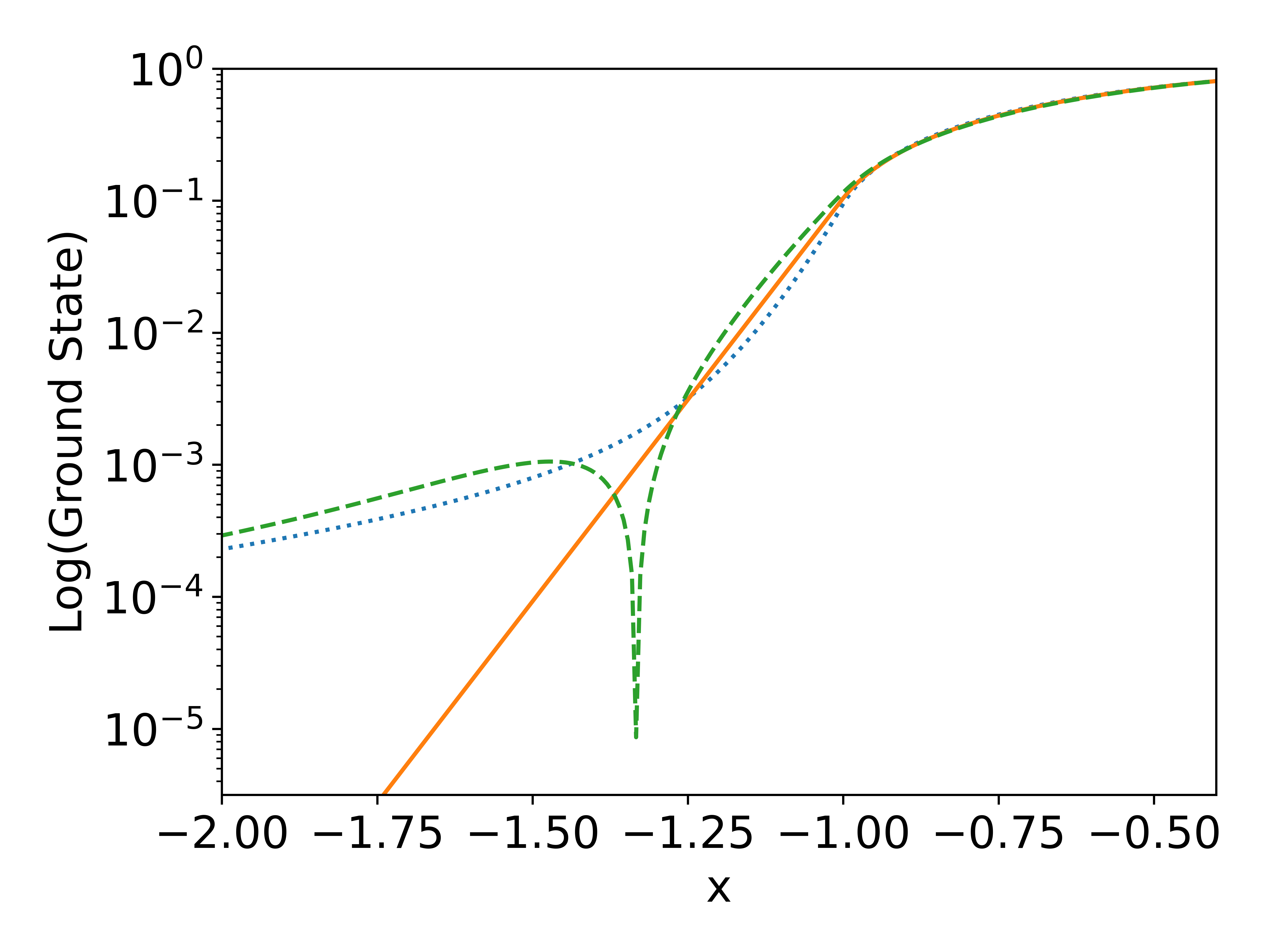}
        \caption{\emph{}}
        \label{fig:sub3}
    \end{subfigure}
    \caption{\emph{Same analysis as in Fig. \ref{fig:QHOPlots}.}}
    \label{fig:test}
\end{figure}

Observation of Fig. \ref{fig:finiteWell} shows that the ground state across fractional orders appears to be changed in only minute and insignificant ways. However, once more, the true difference lies within the tails within the classically forbidden region as is seen in Fig. \ref{fig:finiteWellLog}.

We may immediately note again that the tails of these eigenstates are no longer well characterized by exponential decay. They are in fact Mittag-Leffler like decay patterns similar to what was obtained in the fractional quantum harmonic oscillator and what was shown in Fig. \ref{fig:MLBehavior}. Given this drastic change to the tails within the classically forbidden region, we may hypothesize that there is an enhancement to quantum tunneling due to the longer range nature of confined states. 

\vspace{15pt}

Furthermore, we may examine the number of bound states within this system. As is well known, eigenstates with energies less than the height of the finite well are bound states and as such they are effectively quantized. We may then plot how many eigenstates are bound across many orders.

\begin{figure}[H]
    \centering
    \includegraphics[width=0.6\linewidth]{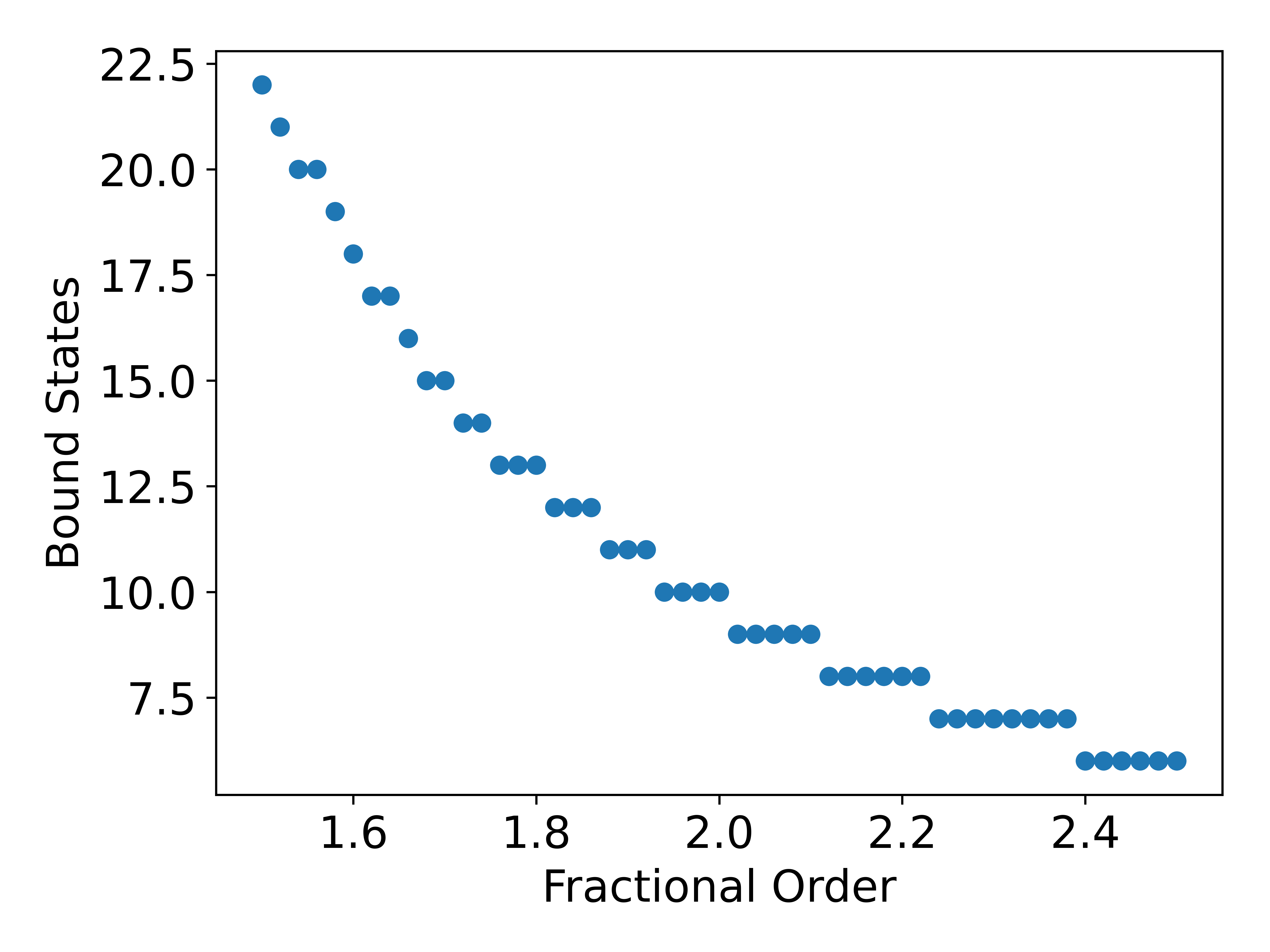}
    \caption{\emph{Number of bound eigenstates in the finite well system across many fractional orders}}
    \label{fig:numberBoundStates}
\end{figure}

It is clear that as the order of the fractional derivative increases so does the energy spacing between states; this results in a smaller number of states that are able to exist within the finite well. Additionally as the fractional order decreases, the number of states that may be placed in the same well increase quite quickly. 

Finally, in the analysis of finite well potentials using our Fourier split-step method with a 6th order Suzuki-Trotter approximation, the adjustment of the time step size plays a critical role in managing the error terms, particularly those associated with the commutator in the Suzuki-Trotter expansion. The non-commutative nature of the kinetic and potential energy operators introduces significant error terms in the presence of discontinuities, as found in steep finite wells. These error terms are inherently linked to the commutator of these operators.

By decreasing the time step size, we directly influence the magnitude of the commutator term in the error expression. In the Suzuki-Trotter approximation, the error is proportional to the higher-order commutators of the Hamiltonian components, which are functions of the time step size. Smaller time steps lead to a reduced impact of these commutator terms, thereby diminishing the overall error. This reduction is particularly crucial in the vicinity of the discontinuities in the finite well, where the standard error terms would otherwise lead to significant inaccuracies.

In the final stages of our simulation, the implementation of an even smaller time step for a brief period serves to further alleviate the distortions caused by these commutator terms. This meticulous calibration of the time step ensures that the evolution of the wavefunction remains accurate, despite the challenges posed by the non-smooth nature of the potential.

\subsection{Double Well}

As a practical example of the enhancement of quantum tunnelling we investigate the quantum double well. A state localized in one well with less energy than the barrier separating them are able to tunnel into the other well through the classically forbidden zone in the middle. We may define the Hamiltonian of this system as the following.

\begin{equation}
    \hat{H} = -\frac{1}{2}\partial_x^{\alpha} + V(x)
\end{equation}

\begin{equation}
    V(x) = -4x^2+\frac{1}{2}x^4+8
\end{equation}

While in the algorithm a large range is chosen as to make sure the eigenfunction is not cut off in the classical forbidden zone we may plot a close up view of the potential below. We also plot the two lowest energy eigenstates which are a symmetric and anti-symmetric wavefunction for order 1.9 as an example. 

\begin{figure}[H]
    \centering
    \includegraphics[width=110mm,scale=1]{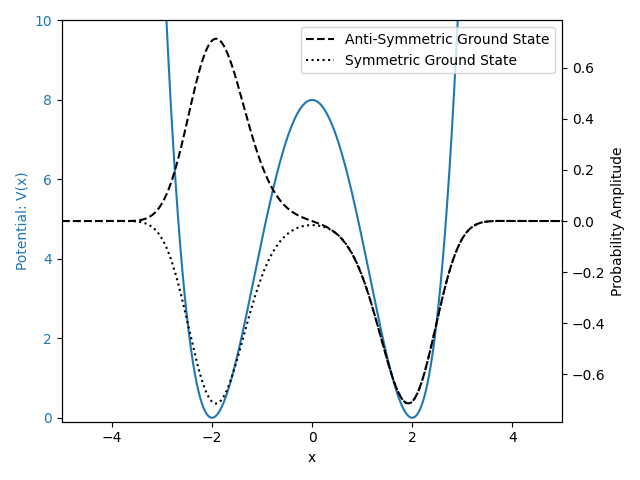}
    \caption{\emph{Potential for double well system along with the two lowest energy eigenstates for fractional order 1.9 as an example}}
    \label{fig:my_label}
\end{figure}

Given these two states, we may put them into an equal superposition that almost entirely isolates it to a specific well. In fact we may construct two superpositions, one in which the wavefunction is in the left well and one that is in the right well. We may write them as the following.

\begin{equation}
    \ket{\psi_L} = \frac{1}{\sqrt{2}}(\ket{\psi_+} - \ket{\psi_-})
\end{equation}

\begin{equation}
    \ket{\psi_R} = \frac{1}{\sqrt{2}}(\ket{\psi_+} + \ket{\psi_-})
\end{equation}

Where $\ket{\psi_+}$ is the symmetric wavefunction and $\ket{\psi_-}$ is the anti-symmetric wavefunction. The reason that tunnelling occurs is that the energy of the symmetric wavefunction is just slightly different from the anti-symmetric wavefunction. Taking the wavefunction in the left well, if we choose to evolve the system in time it will take on the following form.

\begin{equation}
    e^{-i\hat{H}t} \ket{\psi_L} = \frac{1}{\sqrt{2}} (e^{-iE_0t}\ket{\psi_+} - e^{-iE_1t}\ket{\psi_-})
\end{equation}

Removing the unimportant total phase we obtain the following.

\begin{equation}
    e^{-i\hat{H}t} \ket{\psi_L} = \frac{1}{\sqrt{2}} (\ket{\psi_+} - e^{-it (E_1 - E_0)}\ket{\psi_-})
\end{equation}

\begin{equation}
    e^{-i\hat{H}\frac{\pi}{(E_1-E_0)}} \ket{\psi_L} = \ket{\psi_R}
\end{equation}

Clearly after at a time of $t = \frac{\pi}{(E_1-E_0)}$ the wavefunction isolated to the left well becomes isolated to the right well as the relative phase between these two states changes. After the same amount of time it will then return to the left well again. This means that these states will oscillate tunnelling between wells with the following frequency.

\begin{equation}
    f = \frac{|E_1 - E_0|}{2\pi}
\end{equation}

We may now show the differences in energies for the first two eigenstates of the double well system across many fractional orders. 

\begin{figure}[H]
    \centering
    \includegraphics[width=110mm,scale=1]{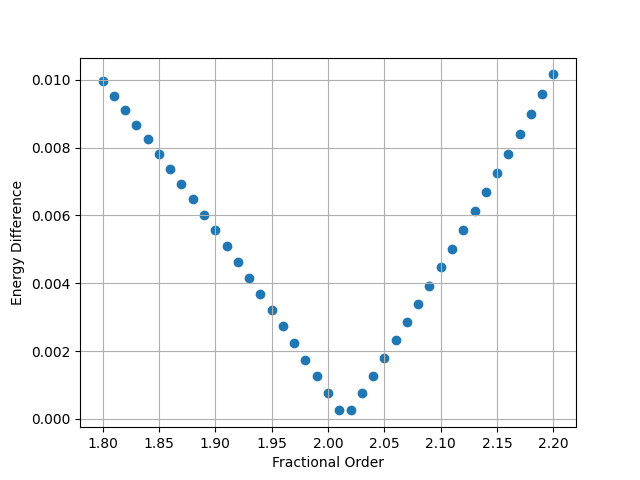}
    \caption{\emph{Energy difference between the two lowest energy states across many fractional orders}}
    \label{fig:my_label}
\end{figure}

As can be seen, the energy difference for this system with a fractional order less than 2.0 is larger and thus in those systems tunnelling occurs at a greater frequency. Interestingly enough when one increases the fractional order at first tunnelling frequency decreases, before rebounding and increasing again similar to the case where one decreased the fractional order. While this plot looks very linear, it is just slightly  not with a small negative second derivative for the data on the order of 10e-6.

\section{Conclusion}

In this article we have constructed an algorithm capable of solving for the eigensystem of the fractional Schrödinger equation with an arbitrary physical potential up to machine precision. Given this tool we have then examined a few physically important systems and discovered that the exponential evanescent waves distinctive to classically forbidden regions are replaced with many of the same properties of the fractional generalization to the exponential function, the \emph{Mittag-Leffler function}. We showed the fractional order of the system can be used to control and enhance tunnelling via the non-locality and enhanced evanescent waves associated with the fractional Schrödinger equation.  We have termed this effect fractionally-enhanced quantum tunneling. This algorithm is a tool that may then be used by experimentalists and applied mathematicians to further investigate the properties of multi-scale materials and discover unique and novel ways to design quantum devices and materials. 

\textit{Acknowledgements} 

We acknowledge useful discussions with Joel Been and Mingzhong Wu. This work was supported in part by the NSF under grant number DMR-2002980.



\bibliography{ref}

\end{document}